# Alignment Theory of Parallel-beam CT Image Reconstruction for Elastic-type Objects using Virtual Focusing Method

**Kyungtaek Jun[a,*], Dongwook Kim[b]**

[a]IM Technology Research Center, 6, Teheran-ro 52-gil, Gangnam-gu, Seoul, 06211, Korea

[b]Department of Mathematics, Atlanta Metropolitan State College, Atlanta, GA, 30310, USA

## Abstract

X-ray tomography has been studied in various fields. Although a great deal of effort has been directed at reconstructing the projection image set from a rigid-type specimen, little attention has been addressed to the reconstruction of projected images from an object showing elastic motion. Here, we present a mathematical solution to reconstruct the projection image set obtained from an object with specific elastic motions: periodically, regularly, and elliptically expanded or contracted specimens. To reconstruct the projection image set from expanded or contracted specimens, we introduce new methods; detection of sample's motion modes, mathematical rescaling pixel values and converting projection angle for a common layer.

## Introduction

X-ray tomography is now considered more than essential in various fields such as biology, archaeology, geoscience and material science[1,2,3,4,5,6] . X-ray tomography is an imaging technique by which the 3D structure of a sample can be reconstructed from 2D projections formed by the penetration of X-rays of different angles. In computer-based image research, the analysis of elastic motion of objects, where through either biological differences or image acquisition structures in the projection image set cannot be achieved without some localized stretching or contracting of the projection image set, has been largely restricted to the study of the motion of rigid-type specimens, where images are assumed to be of specimens that simply need to be rotated and translated with respect to one another to achieve correspondence. It is difficult to reconstruct the projection image set from an object showing elastic motion due to a constraint to some degree of continuity or smoothness; that is, the sample images from elastic objects can be captured in different sizes depending on the angle during the beam time. Mathematical modification is required to produce consistently sized projection image set from the projection images of an object showing elastic motions. The modified projection image has an ideal sinogram pattern within each common layer. This issue has been linked to an alignment solution and thus far, alignment solutions have been largely addressed[2,7,8,9,10,11,12,13].

Recently Jun and Yoon (2017) have proposed alignment solutions for computed tomography (CT) image reconstruction using fixed point and virtual rotation axis[7], which provides a mathematical solution for the CT image reconstruction of a rigid-type object if sample images have errors with translation or tilt. These approaches encourage the reconstruction of projected images from an object showing elastic motions. In this paper, we provide mathematical solutions that can be applied to CT images to reconstruct the image of an object with specific elastic motions; periodic, regular, and elliptical expansion or contraction. We focus on creating a rigid-type object by mathematically modifying a regularly or elliptically moving object in case the image cannot be obtained in rigid-type form. Then, we reconstruct images by rescaling pixel values in CCDs (charge-coupled device) through mathematical correlations to the pixel value of desired sized sample (see Methods) and transforming them into an ideal sinogram pattern using a fixed-point method[7]. The main benefit of our method is that the size of the sample can be adjusted freely in the projection images set obtained from elastic-type objects showing the above motions, that is, the image reconstruction is possible using alignment through a sinogram. We use the reconstruction of the actual image sample as the experimental data and select the sample with the ring artifact to provide a mathematical solution.

**Basic Concepts of Common Layer**

The common layer is a specific layer of an object that is common to all projection images in the projection of the object to obtain information. The projection image for each angle must contain information about the same position of the same object. It is impossible to obtain the correct reconstruction image in the absence of the common layer. That is, to obtain a correct reconstruction image for an object, a common layer is required. The projection image should contain a common layer, and this layer must be secured before any reconstruction work is done. Figure 1 shows the common layer of the sample types that we are addressing in this paper.

The rigid-type specimen in reconstruction only satisfies the Helgason-Ludwig consistency condition. The common layer should contain a common part of the height of the reconstructed object, and for a rigid-type specimen without tilting error, it represents the same axial level of the projected specimen in the projection set. If the shape of the object, however, changes during the beam time, it may contain information about the same layer in some cases, but if it changes in size in the axial direction, it should be considered a different object and information contained in it will also vary. Therefore, to reconstruct a projection image set of a different sized object, it is necessary to convert it to the same sized object; that is, the transformation of the information should precede so that we have the same information as the object we are reconstructing. The projection set for the elastic specimen needs to be converted to the projection set for the rigid-type specimen.

**Determination of a sample's motion modes**

Deformations of elastic objects in mathematics are based on the motion that the shape of an object changes periodically (periodic motion), the motion in which the size is expanded or contracted while maintaining the shape (regular motion) or the motion in which the size is expanded or contracted in a specific direction (elliptical motion). Here, we define motion modes in three categories because it is possible to apply the size changes in various directions through the study of the elastic object which changes in size in a specific direction.

It is possible to determine the motion modes using four fiducial makers (FMs) which are one of fixed points. Figure 2 shows how to determine the motion types by measuring distances among four fiducial makers. Let us there are four fiducial makers on the top, bottom, right and left edge in a sample (Fig. 2a). The distance between top and bottom FMs is denoted by $d_1$ and the other distance between right and left FMs is denoted by $d_2$. For a projection angle ($\theta$), two distances ($d_1$ and $d_2$) in the CCD can be represented by $d_1 * \cos\theta$ and $d_2 * \sin\theta$, respectively (Fig 2b). Similarly, for the contracted image sample, those distances can be represented by $d_1' * \cos\theta$ and $d_2' * \sin\theta$, respectively (Fig 2c and 2e). To determine the motion mode, it is required to measure the distance between FMs in the sinogram where the sample shows elastic motion. Once we decided the desired size of sample at specific projection angle, we could measure $d_1' * \cos\theta$ and $d_2' * \sin\theta$ from the sinogram (Fig 2d and 2f). The conditions for determining motion modes are as follows

$$\frac{d_1'}{d_1} = k\frac{d_2'}{d_2} = \begin{cases} 1 - \alpha(\theta), when\ d_1 > d_1' \\ 1 + \beta(\theta), when\ d_1' > d_1 \end{cases}$$

, where $k$ is a motion mode coefficient. $\alpha(\theta)$ is a coefficient of contraction, $\beta(\theta)$ is a coefficient of expansion.

The elastic-type sample shows the regular motion (contraction or expansion) if $k = 1$ and the elliptical motion if $k \neq 1$.

If the sample conserves the linearity, the projected center of attenuations can be used as fixed point[7]. For this case, two fixed points are enough to determine the mode of sample motion since $d_1$ and $d_2$ can be measured through the distance between the projected center of attenuation and the two fixed points. It is also possible to use three fixed points to find the distance numerically. However, when four fixed points are on the axis, the motion modes can be calculated more accurately.

**Periodic Motion**

In special cases, the shape change of an object may have periodicity, for example, blood vessels that grow and contract in response to the heartbeat and movements of the lung as it regularly expands and contracts. In these cases, the shape of an object that we reconstruct will appear periodical. If the periodicity is constant and the cycle can be recognized, a common layer can easily be obtained. For example, if a heart rate is constant at 120 beats per minute, we will obtain the same effect as taking a frozen heart or vein up to twice a projection every 0.5 seconds when we obtain information about the heart or its associated blood vessels. If the motion is fast for an X-ray dose, it is effective to take one image near maximum or minimum with little change at every period of the periodic function. The constant change of projection angle is the most efficient but not necessarily equal. If the elastic specimen that we want to reconstruction is in contact with other elastic specimens and is periodically affected by each adjacent specimen, we can obtain a projection of the rigid-type specimen in the same way in all the specimens' common cycles.

**Objects that Regularly Expand or Contract**

First, we study a case in which an object is contracted (or expanded) regularly but not periodically in the common layer during a scanning time. In this case, the shape of an object changes circularly or regularly at the same ratio. It can also be used even when the shape of an object is unknown because it is hard to find the periodicity. It is very difficult to predict the result that the size of the object changes during the beam time because the shape changes depending on the angle at which the projection is obtained. If there are two or more fixed points that contain a fiducial marker, we can use them to calculate how much the shape of the specimen has changed[7,14,15,16,17,18,19].

Let us consider the case in which the size of an object regularly contracts with the same ratio. Figure 3 (a) shows three different sized samples which are captured at different angle (0°, 90°, 180°) during the beam time. The size of object is reduced continuously in scanning time while maintaining the total mass attenuation coefficient[7] (MAC) which is measured by the unit voxel of real specimen. The sinogram in the process of deformation for each projection angle is given in Fig. 3(b). The amplitude of the sinogram is gradually decreasing with respect to the projection angle. In this simulation, we used 1200 projected images with the contraction rate of 0.0007% for each angle. That is, the projection angle changes 0.15 degrees, based on NSLS X2B x-ray beamline at BNL.

However, there is a discontinuity at each projection angle due to the existence of discontinuity in density in two adjacent projection images. This implies that there are translation errors that we need to resolve. To reconstruct image from the projection data, we should understand how the information is contained in each CCD in the process of obtaining the projection. Figure 4(a) and 4(b) show the difference in the modified pixel value of a CCD for both the original image specimen ($\Delta l_1$) and contracted image specimen ($\Delta l_2$) for the common area. The CCD in the common area for the contracted image specimen receives less the modified pixel values of the CCD for the original. However, the information on the CCD of the $\Delta l_1$ is the same as the information on the CCD of the $\Delta l_2$. By understanding this relationship and reflecting it in the sinogram to adjust the size without changing the internal pattern, we could reconstruct the image that we want (Fig. 3 and Supplementary Fig. S1). Therefore, if we convert linearly modified pixel values from original sized projection data into a pixel value in a column of a sinogram obtained from a desired size of image specimen, we can obtain a sinogram obtained from the contracted image (Fig 3d right). Conversely, if we divide one linearly modified pixel values from the contracted image into linearly modified pixel values in a sinogram from the original image, we could obtain a sinogram for the size of original image specimen (Fig 3d left). After rescaling pixel values, it needs to change to ideal sinogram patterns by removing translation error using the virtual focusing method[7] (Fig 3e). Finally, the reconstructs will be obtained using the inverse radon transform (Fig 3f).

## Objects that Elliptically Expand or Contract

If the shape of the object changes elliptically rather than regularly, more consideration is needed. Figure 4(a) and 4(c) show two image specimens. One is the original size image specimen, and the other changes the size elliptically. In this case, we also need to transform the original image specimen to the desired sized image through the modification of the sinogram. In a regular motion of an elastic-type sample, there is no pattern change of sinogram according to the projection angle. However, if the size changes elliptically, projection images at the same angle will not show the same pattern of sinogram as the original one. To obtain the desired sinogram, we first need to know the correlation between the pattern for the angle obtained from the original image and the pattern for the angle obtained from the deformed image. If a beam is projected from an object in a $\theta$ direction and the image is the elliptically contracted, the projected beam in the different angle $\theta'$ direction passes through the same part of the object as the original image, and its pattern is also the same.

Figure 5 shows how to reconstruct projection images showing elliptical motion. The captured projection images at different angles ($0^o$, $90^o$, $180^o$) are shown in Fig. 5a. The sample contracts with the reduction rates 0.0005% in x-direction and 0.00025% in y-direction. A sinogram for elliptical sample's motion is shown in Fig. 5b. To reconstruct a projection image set, it is required to modify the projection image set for this angle using the method of $p$ by $q$ image sample (see Methods). After the modification, the sinogram obtained from the contracted object can be transformed into the sinogram of the original sized specimen (Fig 5d, left). It is possible to adjust the sinogram of a contracted object, and it is also possible to fit the sinogram produced by the original object to the contracted sinogram (Fig 5d, right). In other words, the size of the object can be adjusted to the desired size through the modification of the sinogram. More generally, it is possible to resize the object size in any direction. While obtaining the projection image of an object, we can reconstruct the image to the size we want if we know how much the object is stretched or shrunk, horizontally and vertically.

## Discussion

Through the adjustment of the sinogram, an expanded or contracted reconstruction image of an elastic-type object can be obtained. Objects that change regularly in size require length modification in the sinogram, and objects that change elliptically require length and angle transformation in the sinogram. When modifying the angle, the spacing of the angles may not be equally spaced. This does not cause a significant problem because inverse radon transformation is defined for angles of unequal spacing. However, if the image is expanded too much in one direction and a projection set is obtained for a change in equally spaced projection angles, the projection angle becomes dense at a certain angle, and in this case the image may be slightly blurred by the filter function. More research on the filter function is needed in the future. (see Supplementary Fig. S4)

The reconstruction of the expanded object appears darker or more black than the original image (because of the lower MAC). However, if the projection image has linearity in the absorption of photons and in the thickness of the specimen, a clearer image can be obtained by adjusting the constant of MAC. The resulting image has no change in the pattern before the adjustment or in the shape of the image. This result suggests that when we irradiates the radiation, we can reduce the radiation dose within the digitally adjustable limits[7]; that is, there is no reason to increase the radiation dose of the object more than necessary, and the radiation image can be obtained in a more conservative way. This result seems to be very meaningful, which can reduce the damage if the radiation dose is harmful, especially in the case of cells and humans.

We used image samples with ring artifacts to check the error of our method. In an image with a ring artifact, the pattern in the image specimen remains unchanged. It is difficult to see the ring artifact precisely if the rotation axis is misaligned or not ideally focused in the reconstruction. In the case of regular expansion, contraction, and periodic motion, reconstruction can be done by adjusting the size to the desired size, even if the projection images for various specimens of different sizes are mixed.

In the case of elliptic expansion, contraction, and periodic motion, it is possible to obtain an ideally focused reconstruction by rescaling modified pixel values of the CCD and the projection angles of each projection image for different size specimens (Fig. 5). Using an ideal sinogram and a resized sinogram in a contracted image is within the error range created by the digital image (Fig. 6b). Therefore, the reconstruction created by resizing the projection image is not an image with errors.

It is not easy to obtain the ideally focused reconstruction of living cells or organs because of their movement. It takes great effort to attain a better reconstruction during the beam time because only when the cells are fixed or have a specific shape is the reconstruction ideally focused. However, using the method presented in this paper, it is expected that an ideally focused reconstruction will be obtained without lethal damage or manipulation to living cells or organs.

It has been proposed that the modified pixel value of the projection image should be changed to the value that is linearly proportional to the length of the specimen associated with the MAC. This implies that the center of attenuation (CA) not only plays the role of a fixed point but also has a large effect on the pixel value of the reconstruction. If we use a projection image set that does not have linearity, we will have a different reconstruction from the original specimen, and mathematical correction will not be possible (see Supplementary

Fig. S5). When projecting images of the human body parts such as the heart and lungs, they are affected by high density areas such as bones at certain angles. This is a concern because it can represent images different from the actual image. Our technology needs more effort to apply to human body.

## Methods

By changing the projection image set of the elastic specimen to the projection image set of the rigid-type specimen, we can create an ideal sinogram that satisfies the Helgason-Ludwig consistency condition. In this section, we present a way to obtain an ideally focused reconstruction of two types of elastic models

### Regular motion Algorithm (P by P contraction)

The specimen showing regular elastic motion has no change in its internal pattern, even if it changes in size. In this case, we can change the size of the original projection image to the size of the contracted projection image by rescaling modified pixel values of the CCD and also can make the contracted projection image size in the opposite direction to the original projection image size. Therefore, to achieve ideally focused reconstruction, the following process is required:

**Step 1**: In the regularly contracted image set, we change the projection set from the density to the projection set represented by the MAC

**Step 2**: Because the original projection image is less dense than the contracted projection image, the MACs of the projection image (the column of the sinogram) should be rescaled to the desired reconstruction size for the desired projection angle in the projection image set (see Supplementary Fig. S1)

**Step 3**: Using the fixed point and $T_{r,\varphi}$ (see Supplementary Method), the sinogram obtained from specimens of different sizes is transformed into the ideal sinogram pattern.

**Step 4**: An ideally focused reconstruction can be obtained from the ideal sinogram.

### Elliptical motion Algorithm (P by Q Contraction)

In the case of an elliptical elastic specimen, even if the projection angle is the same, different patterns can be seen among the projection images when the size of the specimen is different. Between the elliptically contracted projection image and the original projection image, there is a different pattern at a specific angle. However, mathematical modification at a particular angle can be used to make the pattern of the elliptically contracted projection image the same as that of the original projection image. After that, the same projection image can be made by rescaling modified pixel values of the CCD. Therefore, to achieve ideally focused reconstruction, the following process is required:

**Step 1**: In the elliptically contracted image set, we change the projection set from the density to the projection set represented by the MAC

**Step 2**: We convert the MACs of the projection image to the desired reconstruction size for the desired projection angle and obtain a new projection angle through the projection angle before transformation. To obtain a new projection angle in the case of a specimen with a $p$ by $q$ contraction from the initial specimen size, the following equation is needed. If the projection angle before the change passes through $(1, m)$ from the origin, then the transformed projection angle passes through $(p, qm)$. By this, we can know a new projection angle $\theta' = tan^{-1}(\frac{qm}{p})$, and how the size is transformed between projection images using two angles can be calculated using the following equation.

$$l_1 = \frac{|m+1|}{\sqrt{m^2+1}}, \quad l_2 = \frac{|p*m'+q|}{\sqrt{(m')^2+1}}$$

where $m' = \frac{q*m}{p}$, $l_1$ and $l_2$ are the distance from a line with the same slope as the projection angle through the origin to $(1, -1)$ and the distance from a line with the same slope as the projection angle through the origin to $(p, -q)$ on the CCDs (Fig. 4c). We rescale modified pixel values of a CCD using the ratio of $\Delta l_1$ and $\Delta l_2$

**Step 3**: Using the fixed point and $T_{r,\varphi}$ (see Supplementary Method), the sinogram obtained from elliptically expanded projection image is transformed into the ideal sinogram pattern.

**Step 4**: An ideally focused reconstruction can be obtained from the ideal sinogram

## Error Calculation

In this section, we investigate the difference in relative errors which are (1) calculated after the center of attenuation[7] (CA) of the resized projection image is transformed to the CA of the original projection image and (2) calculated after the CA of the original size projection image that moved one voxel and the CA of the original size projection image that did not move. The relative error can be calculated by the following equation

$$E(\theta) = \frac{\sum_K ||P1_\theta(k) - P2_\theta(k)||}{N}$$

where $N$ is the total pixel number of the region of interest and $P1_\theta$ and $P2_\theta$ are original projection image and the changed projection image, respectively.

Figure 6 shows the sample image where the gray part is the region of interest (left) and the two relative errors in terms of the angle ($\theta$; right). In the right panel, the blue curve is the relative error of (1) and the red curve is the relative error of (2). There is a difference from the specific angle depending on the image. In the projection image, a peak occurs when the high-density pixel is severely divided by the angle (see Supplementary Fig. S3). In general, relative error has the property that the larger the image, the smaller the relative error. The relative error of the resized density is similar as the maximum value of the relative error, which is different from the actual value of one pixel—relative error continues based on position. If the pixel size is reduced, relative error will also be reduced.

## Acknowledgement

The authors acknowledge Dr. Keith Jones, a senior physicist at Brookhaven National Laboratory, provided the soil projection sets for our image sample.

## Author contributions Statement

K.J conceived and designed the theoretical experiments, and performed the experiment.

The theoretical solutions were developed by all authors, and all authors wrote the paper.

## Competing financial interests

The authors declare no competing financial interests.

# Figures & Figure Legends

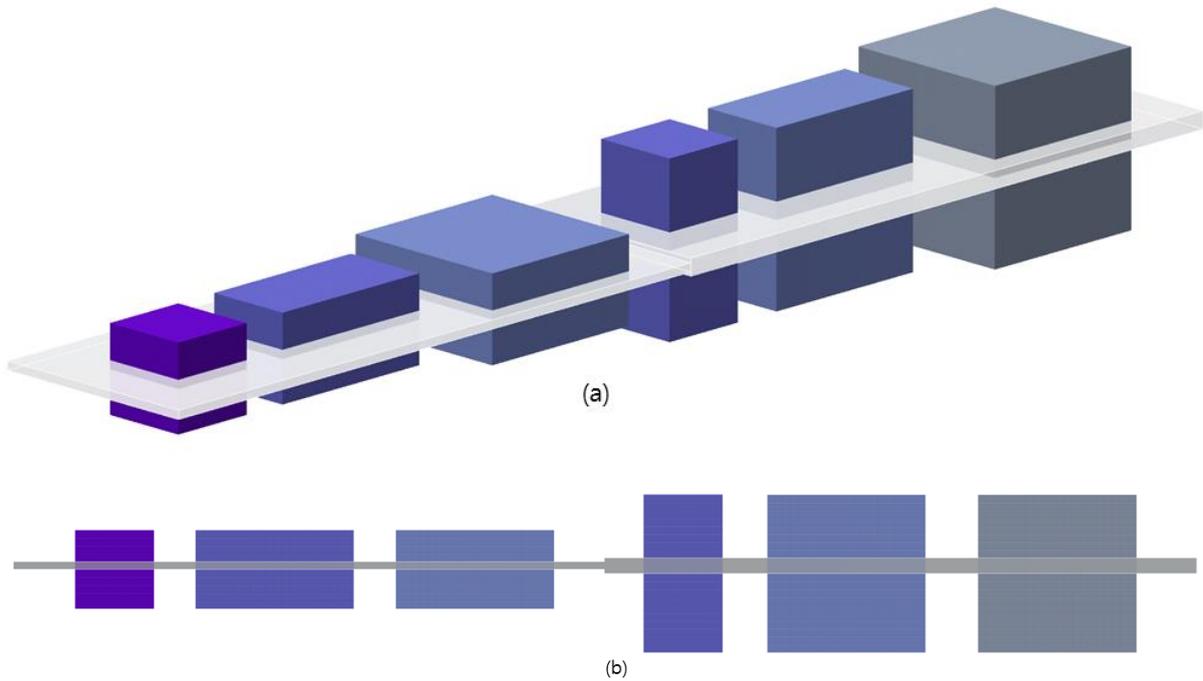

**Figure 1. The change in the common layer when the size and height of one object change.** The first object (purple) has been used to create other five objects by changing the size or the height. The sizes of the first three samples from the left are 10x10x10, 20x10x10, and 20x20x10, respectively. The next three samples are the samples whose heights from the first three samples have been doubled (10x10x20, 20x10x20, and 20x20x20, respectively). When the height of the first three samples is doubled, the height of the common layer also doubles.

(a) 3D image samples in space
(b) 3D image samples in front view. The gray part indicates the common layer

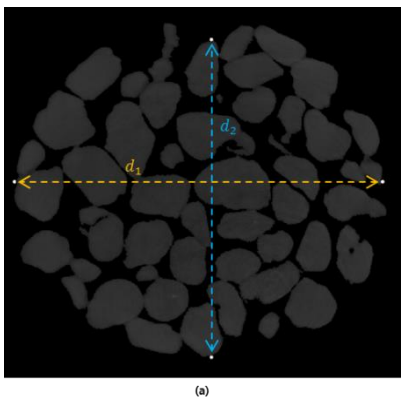

(a)

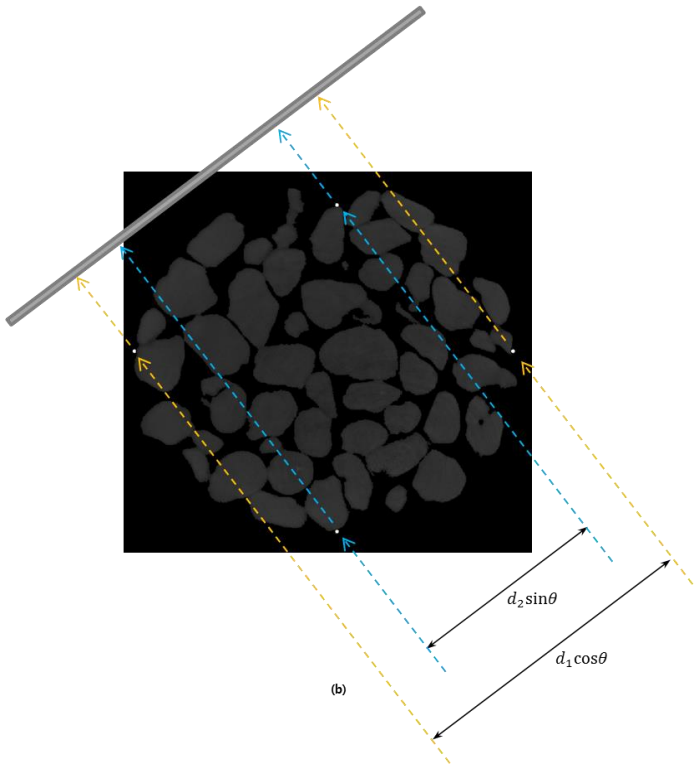

(b)

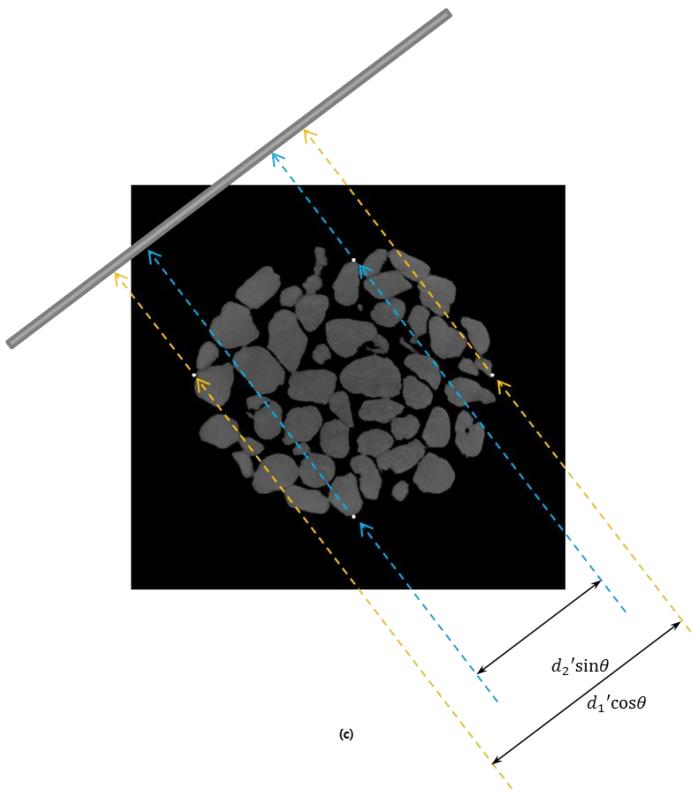

(c)

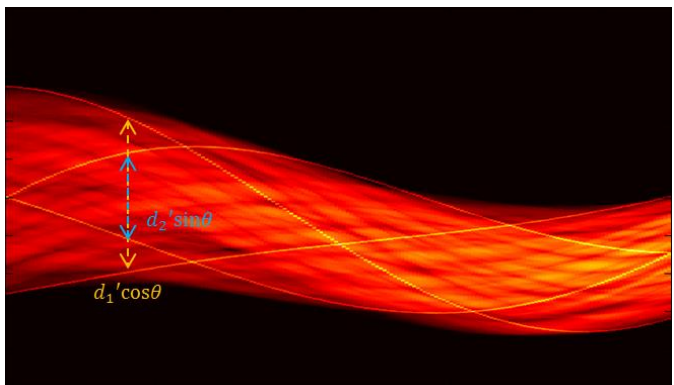

(d)

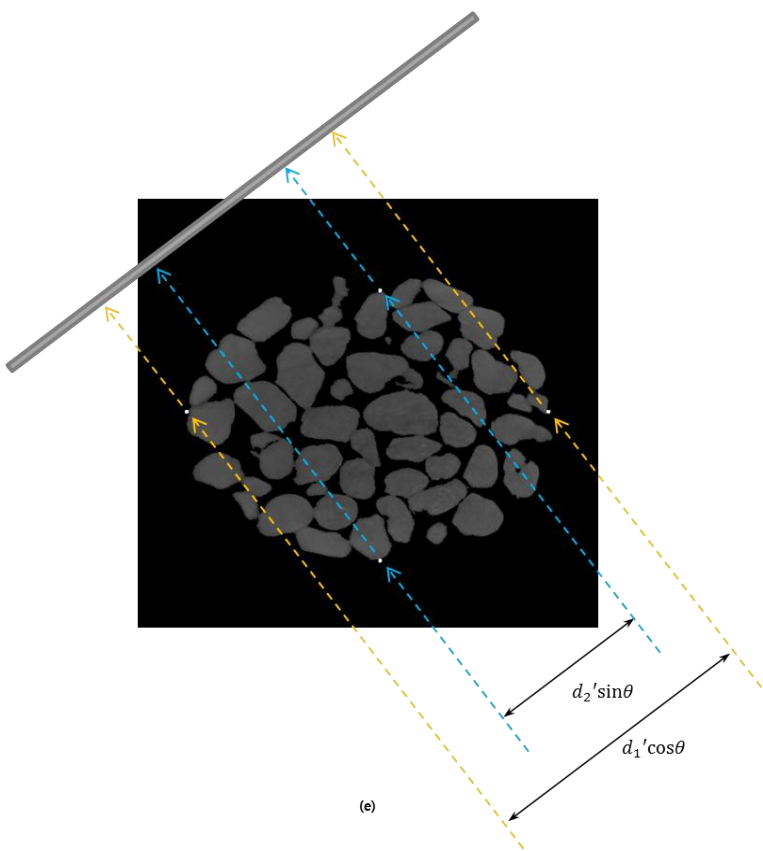

(e)

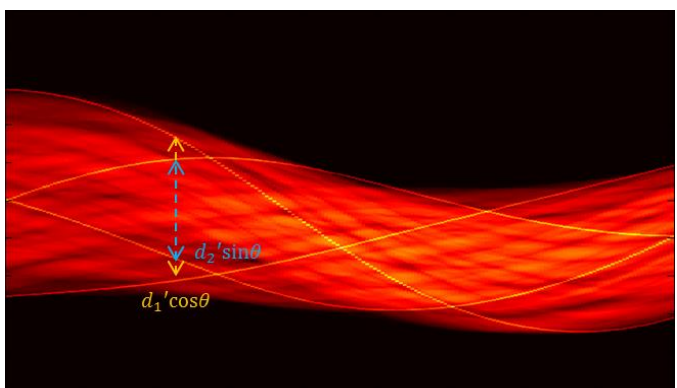

(f)

**Figure 2. Determination of motion modes using four fiducial makers**

(a) Original image sample with four fiducial makers (high density area); $d_1$ is the distance between top and bottom fiducial makers and $d_2$ is the distance between right and left fiducial makers

(b) The distance between fiducial makers on the original image in the CCD; $\theta$ represents a projection angle.

(c) The distance between fiducial makers on the regularly contracted image in the CCD.

(d) The sinogram in the process of regular contraction. High density in the sinogram represents the circular trajectory of fiducial maker.

(e) The distance between fiducial makers on the elliptically contracted image in the CCD.

(d) The sinogram in the process of elliptic contraction. High density in the sinogram represents the circular trajectory of fiducial maker.

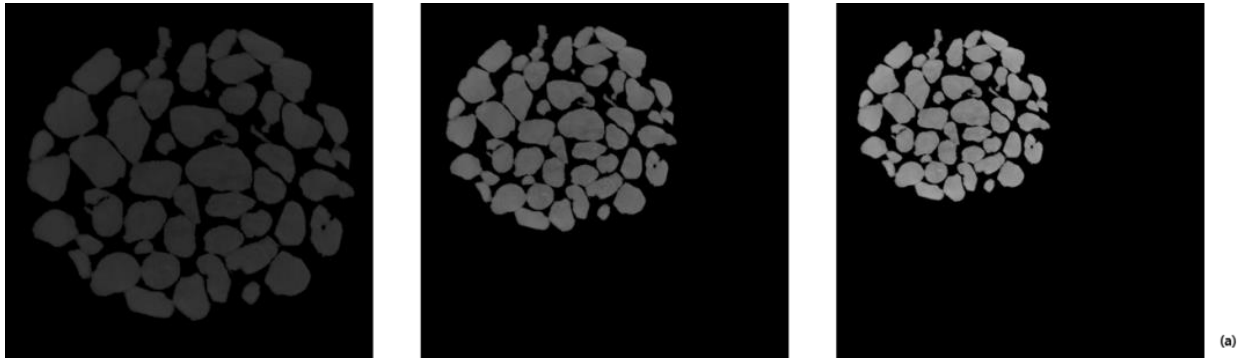

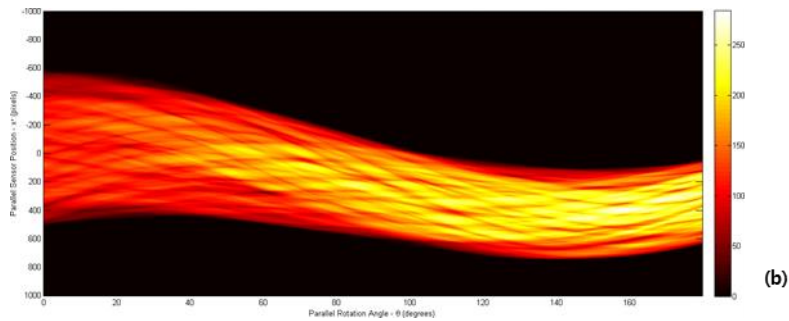

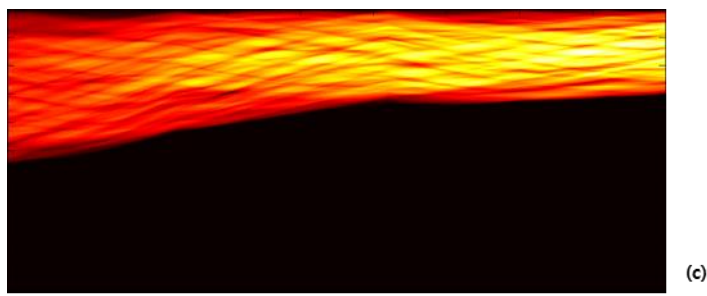

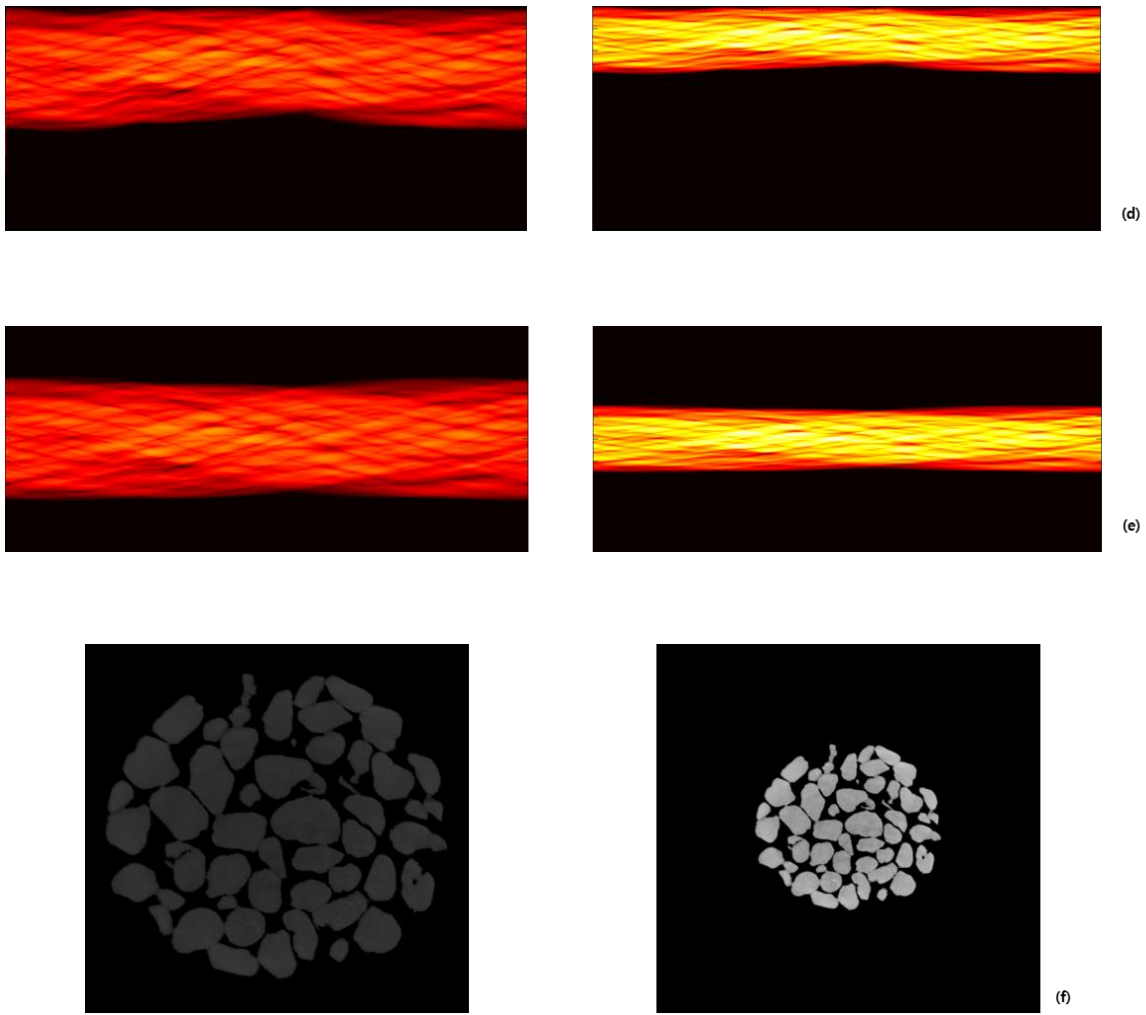

**Figure 3. The reconstruction of an elastic-type object showing regular contraction continuously in time**

**(a)** The process in deformation of elastic-type object during a beam time. The different sized projection images are captured at different angle (0°, 90°, 180°)

(b) The sinogram in the process of deformation. The amplitude of the sinogram is gradually decreasing with respect to the projection angle.

(c) The modified sinogram by moving the projected shadow of the object to rescale the pixel values

(d) The sinogram after rescaling pixel values; rescaled to contracted final sized sample (right) and rescaled to original sized sample (left)

(e) Changed the ideal sinograms from each sinogram in (d) with translation error using projected fixed point method

(f) Ideally focused reconstructions from the sinograms (e), The reconstructions with final sized contraction (right) and original sized (left)

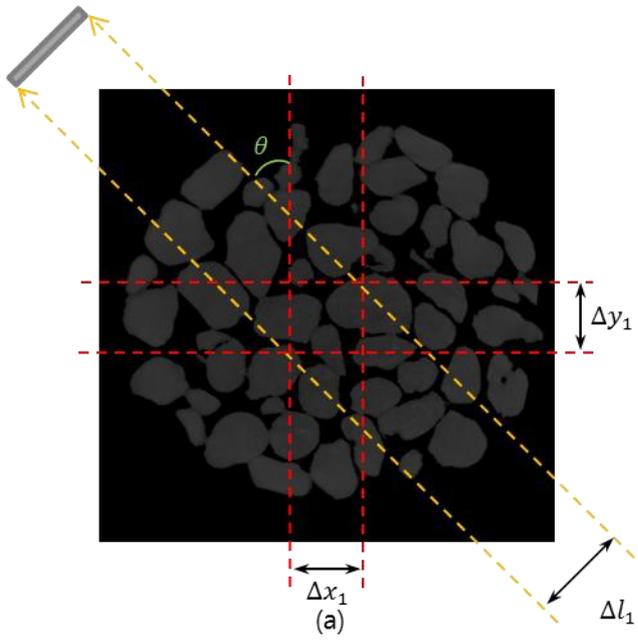

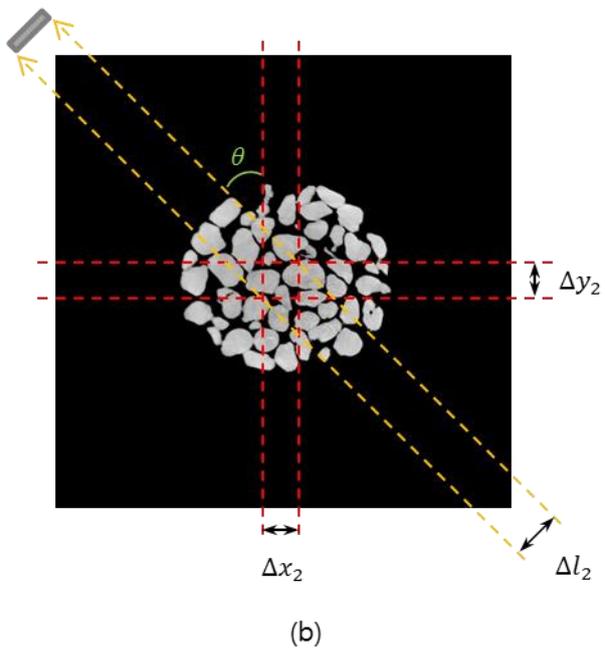

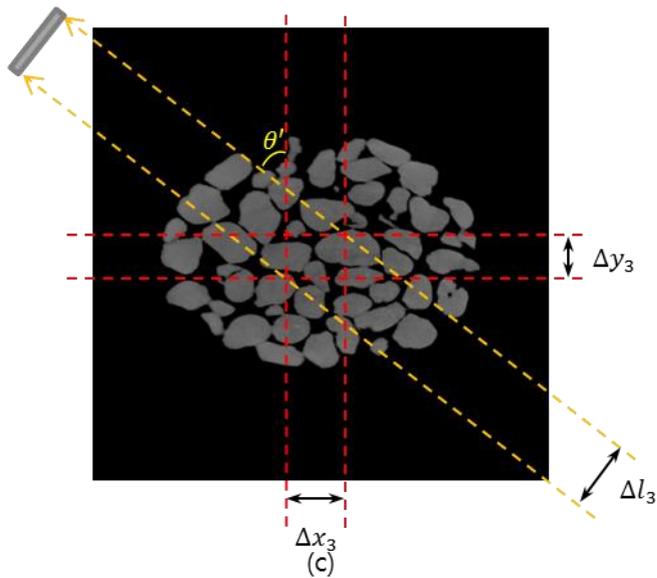

**Figure 4. When the X-ray penetrates samples of different sizes, the changes both in the CCD and in the angle for the common area of the sample**

(a) Original image specimen with a projection angle: When the X-ray for projection angle of $\theta$ passes the crossed area of the image sample, the size of the area projected onto the CCD is $\Delta l_1$

(b) Regularly contracted image specimen; The X-ray angle for the common area does not change by $\theta$, but the size of the area projected onto the CCD decreases to $\Delta l_2$

(c) Elliptically contracted image specimen; the X-ray angle for the common area changes to $\theta'$ and the size of the area projected onto the CCD is changed to $\Delta l_3$

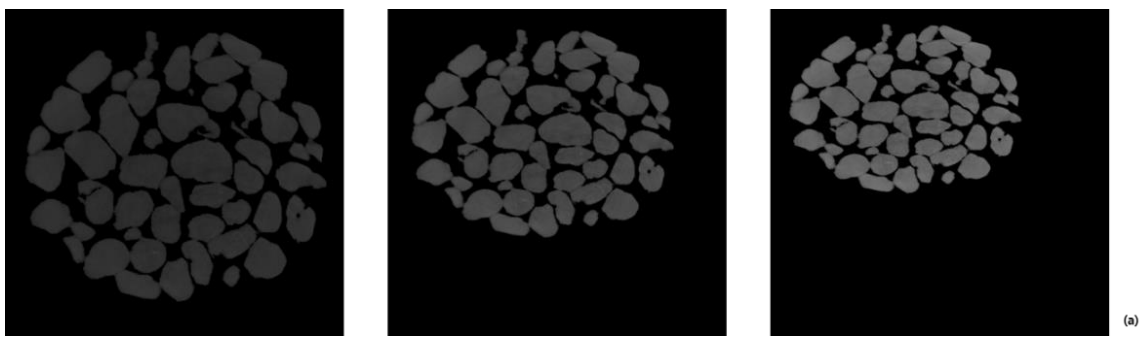

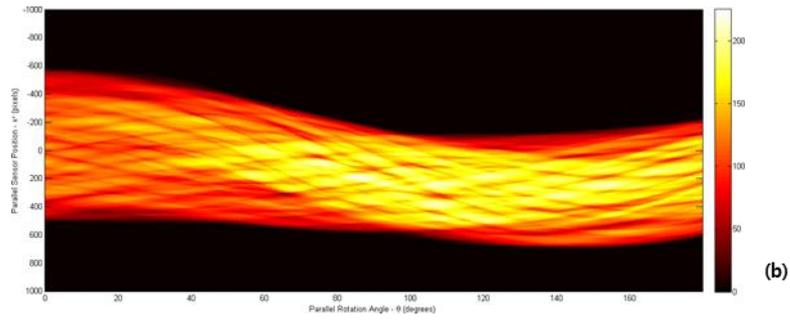

(b)

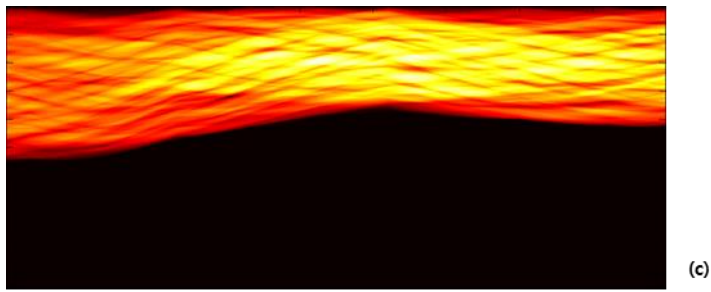

(c)

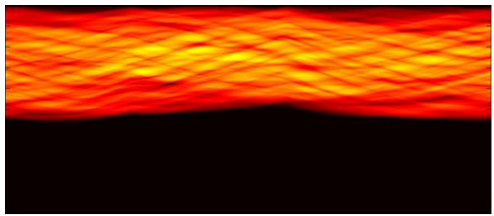 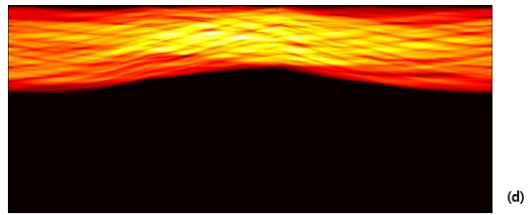

(d)

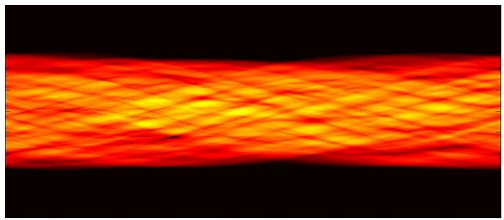 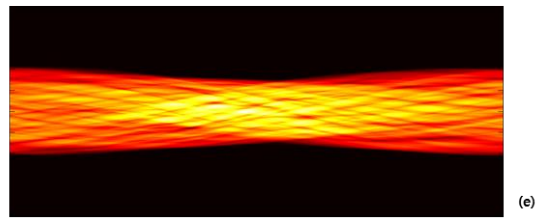

(e)

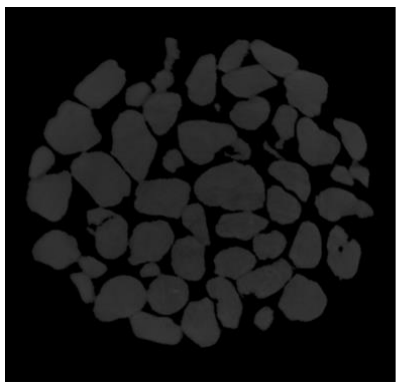 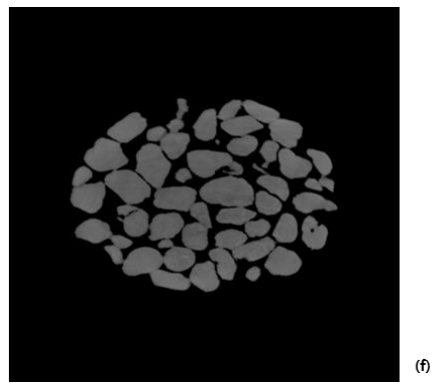

(f)

**Figure 5. The reconstruction of an elastic-type object showing elliptical contraction continuously in time**

**(a)** The process in deformation of elastic-type object during a beam time. The different sized projection images are captured at different angle (0°, 90°, 180°)

(b) The sinogram in the process of deformation.

(c) The modified sinogram by moving the projected shadow of the object to rescale the pixel values

(d) The sinogram after rescaling pixel values; rescaled to contracted final sized sample (right) and rescaled to original sized sample (left)

(e) Changed the ideal sinograms from each sinogram in (d) with translation error using projected fixed point method

(f) Ideally focused reconstructions from the sinograms (e), The reconstructions for final sized contraction (right) and original sized samples (left)

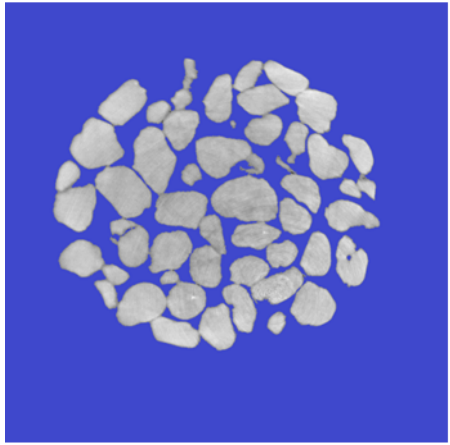
(a)

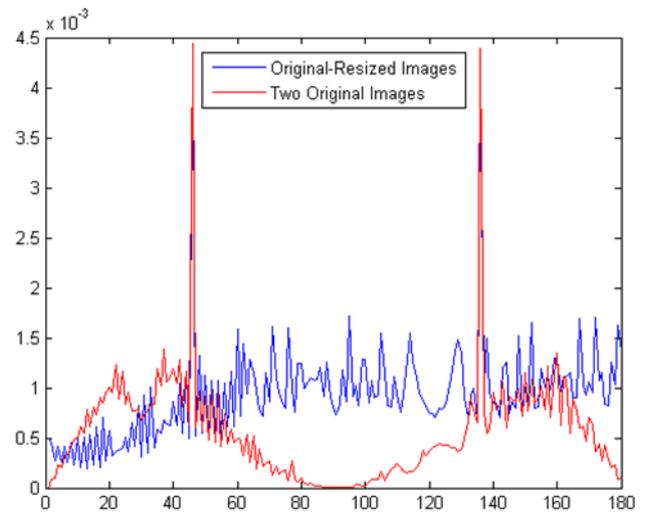
(b)

**Figure 6. The difference in relative errors.**

(a) The 746x746 image sample where the gray part is the region of interest
(b) The blue curve is the relative error that is calculated after the center of attenuation[7] (CA) of the resized projection image is transformed to the CA of the original projection image. The red curve is the relative error that is calculated after the CA of the original size projection image that moved one voxel and the CA of the original size projection image that did not move

# Alignment Theory of Parallel-beam CT Image Reconstruction for Elastic-type Objects using Virtual Focusing Method

**Kyungtaek Jun[a,*], Dongwook Kim[b]**

**Supplementary Figures**

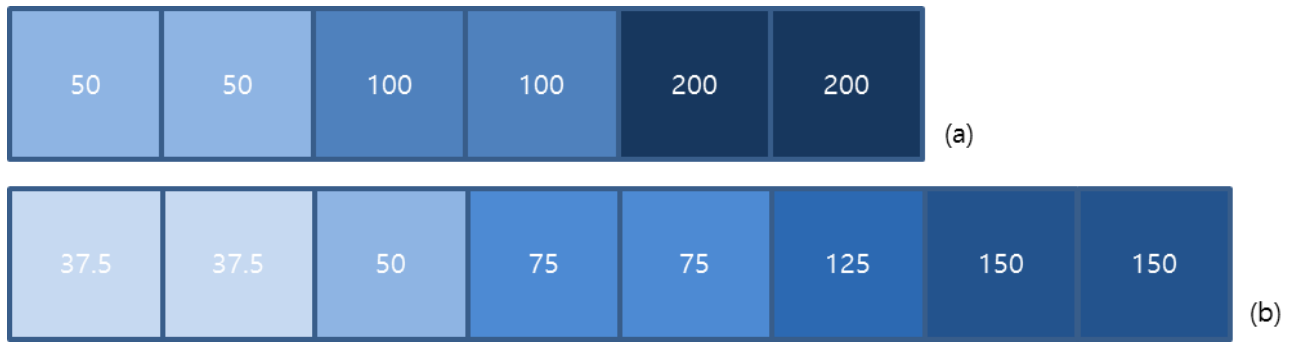

**Figures S1. Rescaling Image with the linear interpolation:**
Resizing pixels 3 to 4 with the same total attenuations. In this case, the third value of the expanded pixel value $P2_\theta(3)$ calculated by $P1_\theta(2) * \frac{2}{4} + P1_\theta(3) * \frac{1}{4}$

(a) Original projection image $P1_\theta$ with pixel values

(b) Expanded projection image $P2_\theta$ with rescaled pixel values

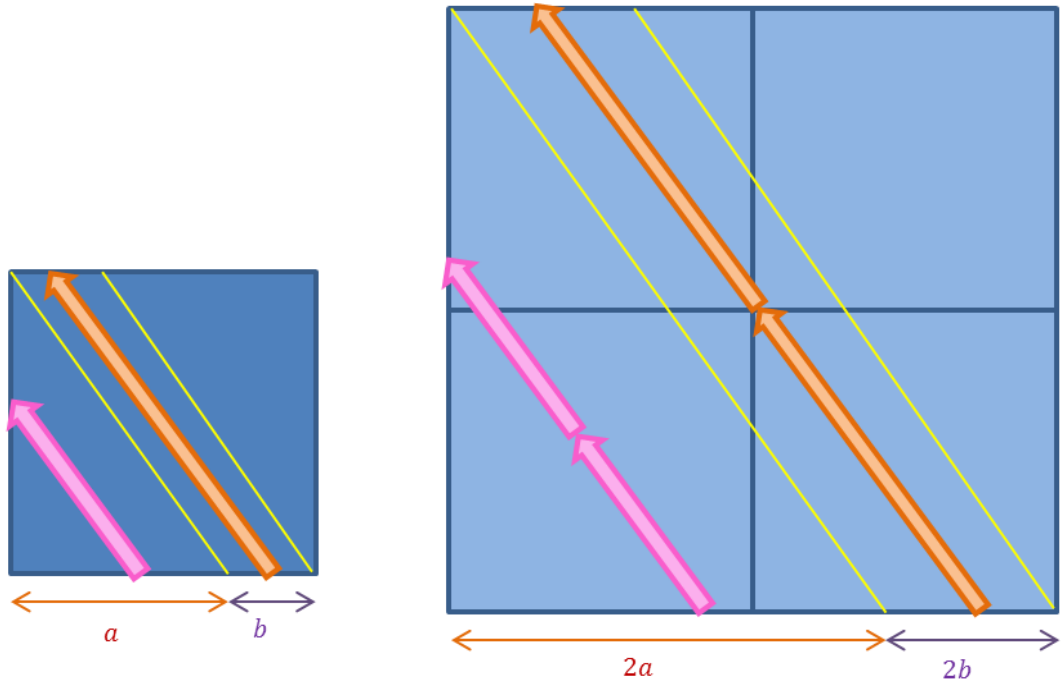

| Object | Original | Doubled xy-direction |
|---|---|---|
| MAC per a voxel | $\alpha$ | $\frac{1}{4}\alpha$ |
| X-ray Path Length $\Delta l$ | $\Delta l$ | $2\Delta l$ |

(b)

**Figure S2**. Basic description for the expanded voxel from the original voxel

(a) An example that the regularly doubled expanded voxels. Each bolded arrow indicate X-ray penetration in $s_1$ and $s_2$ sections and their region will be doubled

(b) Basic information table

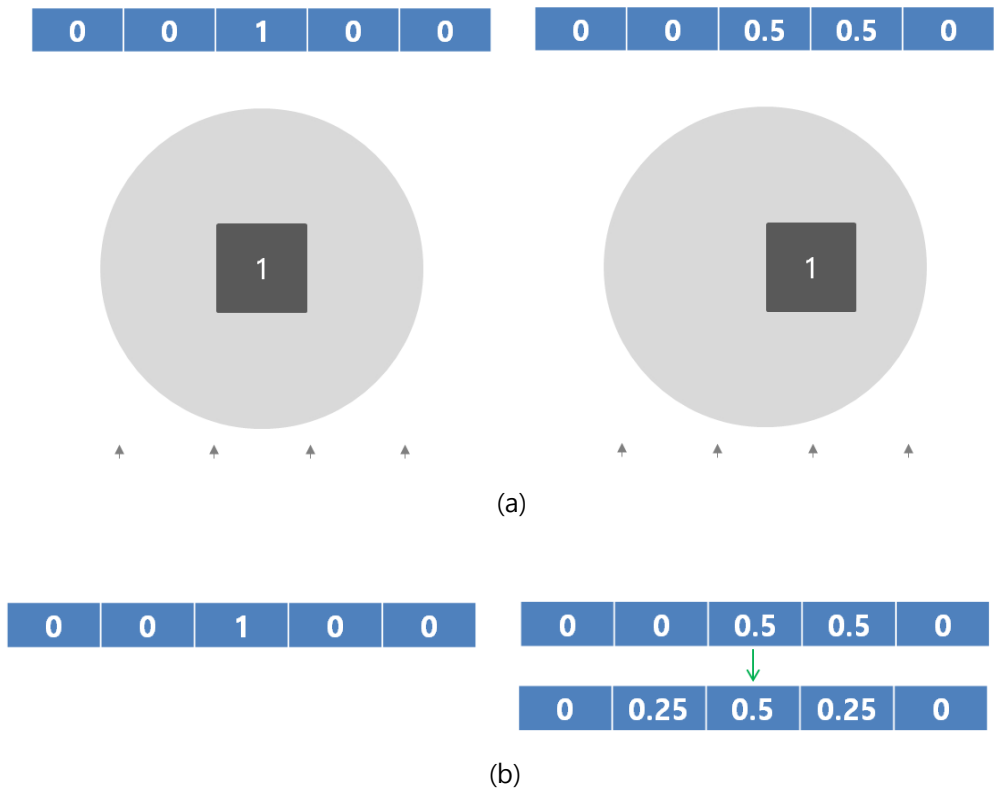

**Figure S3. An example when the worst case of relative error occurs because of the limitation of the digital image**

(a) A specimen on the stage with two different positions. The specimen is placed on the center of the stage (left panel). The specimen moves from the center to the right 0.5 voxel (right panel).

(b) Translate the right projection image which has the same projected CA position of the first projection image in (a). In this case, the relative error is 1.

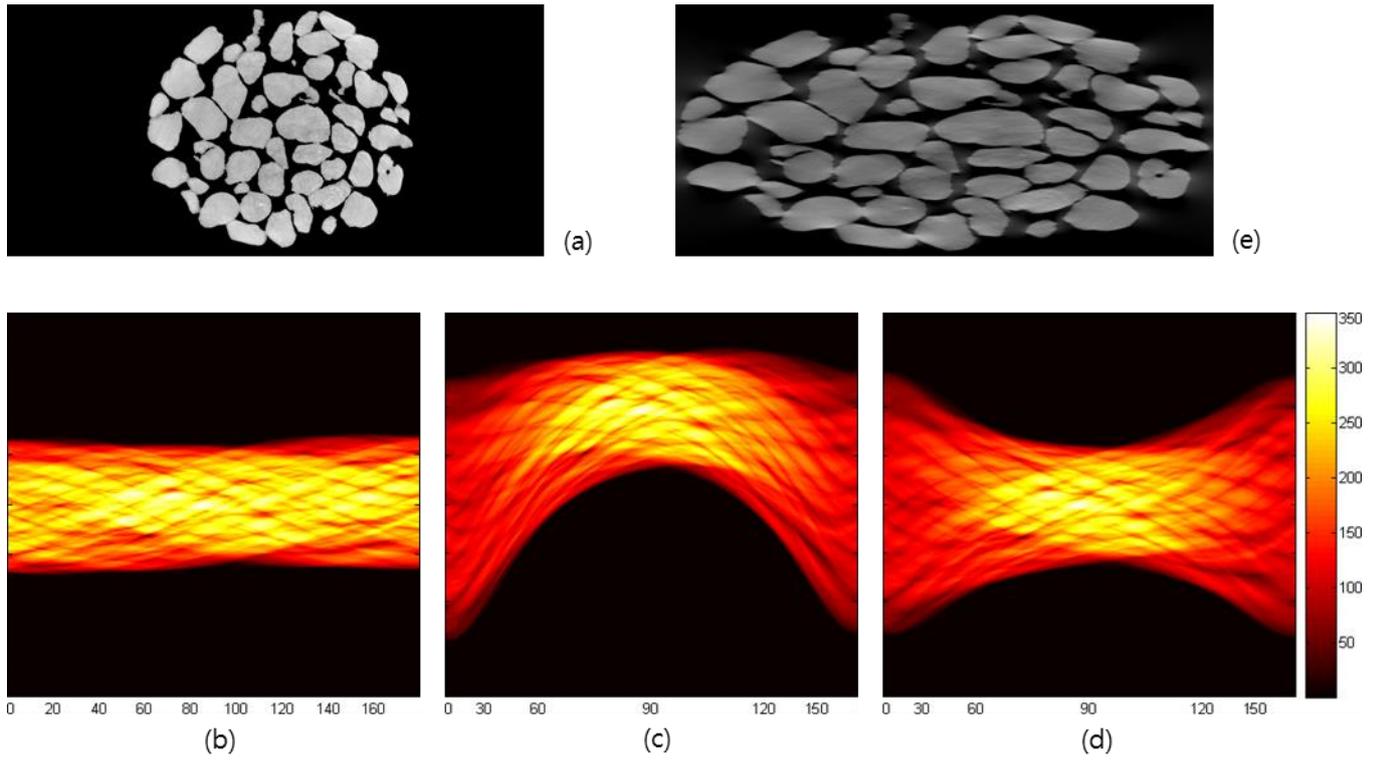

**Figure S4. The elliptically expanded reconstruction from the original projection image set.** If the ideal sinogram with evenly changed projection angle is modified to a sinogram of an image that is severely expanded in x-axis direction, the ideally focused reconstruction is not clear.

(a) A specimen, (b) Ideal sinogram of (a), (c) A sinogram that is elliptically doubled in the x-axis direction from the sinogram (b), (d) Translate each $\overrightarrow{P_{CA}}$ in columns of the sinogram onto $T_{0,\varphi}$, (e) Ideally focused reconstruction obtained from (d)

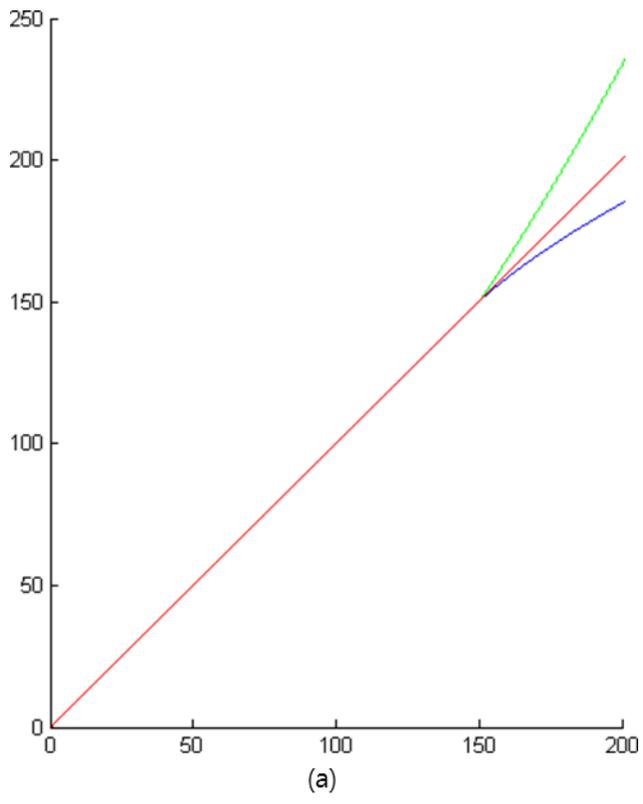

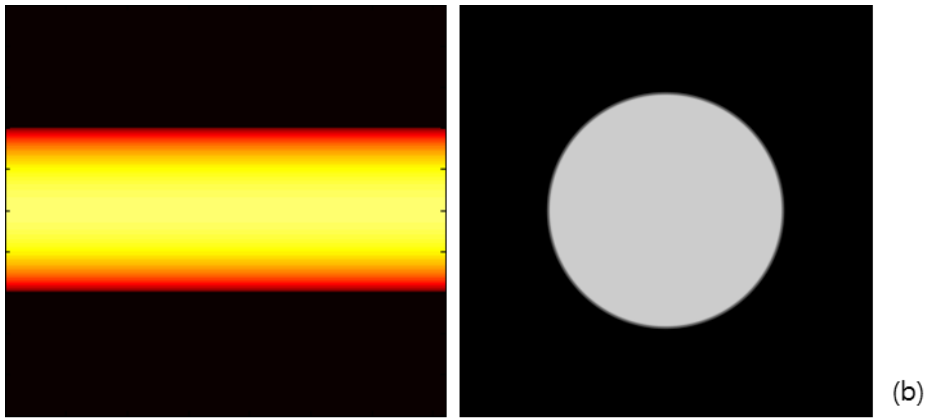

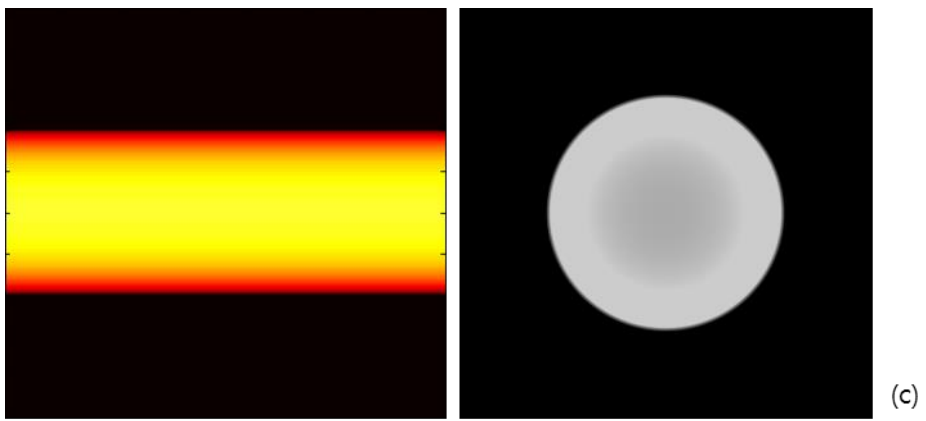

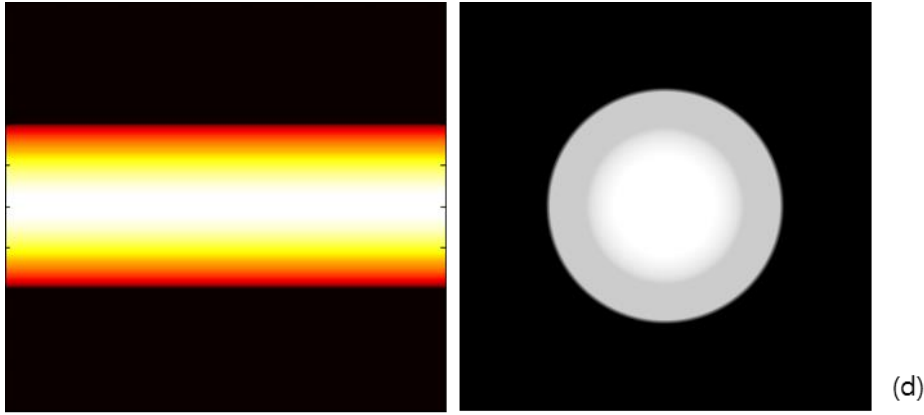

**Figure S5. For a circle with the radius 100, the change in reconstruction according to the relationship between X-ray path length and density.**

(a) Three types of density functions related to X-ray path length. The red line shows linear relationship

(b) A sinogram and its reconstruction using the linear relationship from the red line in (a)

(c) A sinogram and its reconstruction using the relationship from the blue curve in (a)

(d) A sinogram and its reconstruction using the relationship from the green curve in (a)

## Supplementary Method

### Virtual Focusing Method using Fixed Point

The circular trajectory of a point $p$ in the real space corresponds to a curve drawn by the sinusoidal function in the sinogram. The function is given by [cite Jun]

$$T_{r,\varphi}(\theta) = r * \cos(\theta - \varphi), 0 \leq \theta < 180° \quad (1)$$

Where $r$ is the distance between the rotation axis and the point $p$. $\theta$ is the projecting angle, and $\varphi$ is the angle between the line $\overleftrightarrow{Op}$ and the orthogonal line to the projection angle at $\theta = 0$.

In ideal cases, the center $O$ is converted to $T_{0,\varphi}$ in the sinogram, but not in the actual sinogram. $T_{r,\varphi}$ is a function that shows how a specific point $p$ in the real space moves on the sinogram. In other words, if a point $p$ in the solid specimen rotates on the stage and the projected curve drawn by the movement of $p$ for each angle is the same as the sinusoidal curve by $T_{r,\varphi}$ in the sinogram, then the projected trajectories of other points in the specimen should satisfy the projected curves by $T_{r_n,\varphi_n}$.